# BC4LLM: Trusted Artificial Intelligence When Blockchain Meets Large Language Models

**Haoxiang Luo[1, 2*], Jian Luo[2], Athanasios V. Vasilakos[3]**

[1] Key Laboratory of Optical Fiber Sensing and Communications (Ministry of Education), University of Electronic Science and Technology of China, Chengdu, 611731, China

[2] Hainan Chain Shield Technology Co., LTD, Haikou, 570311, China

[3] Center for AI Research, University of Agder, 4879 Grimstad, Norway

**Abstract**

In recent years, artificial intelligence (AI) and machine learning (ML) are reshaping society's production methods and productivity, and also changing the paradigm of scientific research. Among them, the AI language model represented by ChatGPT has made great progress. Such large language models (LLMs) serve people in the form of AI-generated content (AIGC) and are widely used in consulting, healthcare, and education. However, it is difficult to guarantee the authenticity and reliability of AIGC learning data. In addition, there are also hidden dangers of privacy disclosure in distributed AI training. Moreover, the content generated by LLMs is difficult to identify and trace, and it is difficult to cross-platform mutual recognition. The above information security issues in the coming era of AI powered by LLMs will be infinitely amplified and affect everyone's life. Therefore, we consider empowering LLMs using blockchain technology with superior security features to propose a vision for trusted AI. This paper mainly introduces the motivation and technical route of blockchain for LLM (BC4LLM), including reliable learning corpus, secure training process, and identifiable generated content. Meanwhile, this paper also reviews the potential applications and future challenges, especially in the frontier communication networks field, including network resource allocation, dynamic spectrum sharing, and semantic communication. Based on the above work combined and the prospect of blockchain and LLMs, it is expected to help the early realization of trusted AI and provide guidance for the academic community.

**Keywords**

Trusted AI, blockchain, large language model, AI-generated content, ChatGPT.

*Corresponding author.

E-mail address: lhx991115@163.com (H. Luo)

## 1. Introduction

The rapid development of AI technology represented by machine learning (ML) and natural language processing (NLP) has produced many representative LLMs [1-3], such as ChatGPT (OpenAI), ERNIE Bot (Baidu), Qwen-7B (Alibaba), and so on. These models can generate multimodal content across different domains, such as text, images, music, etc., hence the name AIGC [4-7]. The class of model builds new data by learning, analyzing, and understanding existing data, and outputs it in a way that is easy for humans to understand. At present, this kind of AIGC model has been widely used in various fields, such as education [8-10], public health [11], scientific publishing [12], even military [13], and so on.

The large language model is based on the Transformer architecture described in [14], which overcomes some of the limitations of previous sequence-to-sequence methods used for NLP, such as recurrent neural networks (RNN) and convolutional neural networks (CNN) [15]. As ChatGPT describes, it trains on a large corpus of text data and fine-tuns the generated content to meet the needs of a specific task. These steps allow ChatGPT to generate content similar to human habits and interact with humans for multiple rounds [16-17]. It is precisely because of the above user-friendly advantages that this kind of LLM is about to move towards a wider range of application scenarios.

However, because LLMs need to learn a lot of corpuses, and the generated content is widely used, once the source of the learned corpus is unknown or even incorrect, the generated content will be unsafe and unreliable. In addition, the ownership of the generated content and the traceability of sharing cannot be guaranteed, thus restricting the use value of such content. Finally, to improve its work efficiency during the training process, the AI LLM may conduct distributed training across multiple data centers, which also has serious hidden dangers of privacy disclosure. Therefore, there are security problems in the **(1) learning corpus**, **(2) training process**, and **(3) generated content** of the LLM, which need to be solved before it is widely used in various industries.

Fortunately, blockchain [18-19], which integrates advanced technologies such as consensus mechanisms, cryptographic algorithms, and distributed databases, offers a promising solution to the above security problems of LLMs, paving the way for LLMs to empower various industries. Recently, because blockchain can ensure the security of distributed systems, and can make nodes in the network reach consistency without a trusted third party, it has been widely used in the Internet of Things (IoT) [20], Internet of Vehicles (IoV) [21-22], Internet of Energy (IoE) [23-24], Industrial Internet of Things (IIoT) [25], decentralized autonomous organizations (DAO) [26], and other fields.

Moreover, blockchain is also being applied to various AI models [27], such as federated learning (FL) [28]. The above examples give us sufficient motivation to introduce blockchain technology to solve the whole process security problems of learning, training, and generation of the LLM, and help its better promotion and become a trusted AI. Contributions in this paper are as follows

- We have introduced blockchain, the LLM, and blockchain-enabled AI in turn, so that readers can further understand related technologies of blockchain and the LLM.
- We have sorted out the security issues that may exist in LLMs, such as learning corpus, training process, and generated content, and analyzed why LLMs need blockchain.
- We have put forward the concept of BC4LLM and analyzed how blockchain technology can empower LLMs, making the above three parts more secure and able to be further applied. To our best knowledge, this is one of the first efforts to propose blockchain for LLM.
- We have listed the possible applications of the BC4LLM. This type of application should have a high demand for security, as well as strict requirements for processing intelligence and data accuracy.
- We have further pointed out some technical challenges and future development directions of BC4LLMs, providing a valuable reference for the academic community's subsequent research.

The remaining structure of this paper is as follows. Section 2 provides background knowledge on blockchain, the LLM, and blockchain-enabled AI technology. Section 3 explains why LLMs need blockchain. Section 4 gives the specific scheme of blockchain empowering LLMs. Sections 5 and 6 look at the potential applications and the possible challenges of the BC4LLM, respectively. Finally, Section 7 summarizes this paper.

## 2. Background knowledge

This section provides background knowledge and related technologies on blockchain, LLMs, and blockchain-enabled AI.

2.1 Blockchain technology

As the backbone technology for many other distributed scenarios such as digital currencies [29-30] and distributed networks [31-32], blockchain has become a revolutionary decentralized data management and storage framework that can build consensus and protocols in a trustless and distributed environment [33].

Essentially, blockchain acts as a decentralized ledger management system that is used to record and verify transactions [18]. It allows participants to reach consistency and complete transactions in a peer-to-peer (P2P) network [34] without authority or a trusted third party. With the help of blockchain, all peers jointly maintain a public ledger for trust, decentralization, security, traceability, transparency, and immutability. The transactions recorded in the blockchain system may be any form of data that involves the share, transfer, or exchange of ownership or resources, such as digital currency, copyright, digital files, non-fungible tokens (NFT), and so on [35-36].

Fundamentally, blockchain is built on a physical network that combines communication, computing, and storage functions, as shown in Fig. 1. The network is the basis for blockchain features such as the blockchain consensus mechanism. Therefore, the blockchain system can be described as a two-tier structure containing both the infrastructure and the blockchain [18]. The infrastructure layer is at the bottom and is responsible for maintaining the P2P network, establishing connections through wired or wireless communications, and simultaneously computing and storing data. In addition, the blockchain layer sits at the top, enabling trust and security in information communication through three entities: transactions, blocks, and chain of blocks [37]. Particularly, transactions contain information necessary for the client and are recorded in the public ledger; Blocks are used as entities that reliably record transaction amounts and other required information; The chain of blocks consists of all blocks linked in an orderly manner to form a chain, which indicates the logical relationship between these blocks.

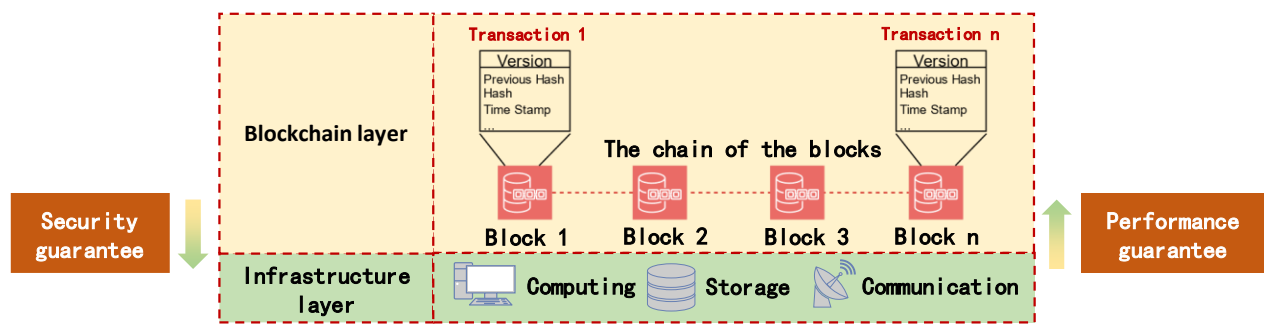

Fig 1. Blockchain network structure.

As a core function of blockchain, the consensus mechanism and smart contract play an important role in ensuring transaction completion and security. Consensus works at the blockchain layer, ensuring the clarity of the transaction order and the consistency of all participants [38]. For the smart contract, it is often triggered when a transaction is reached and the blockchain is updated. The execution process is completely automated [39]. Therefore, the smart contract is often seen as a function that guarantees the automation and authority of the blockchain system operation [40-41].

It is precisely because of the above security features of blockchain that it is considered to be an important enabling technology to make AI reliable and trusted.

2.2 Large language model

The LLM assigns probabilities to sequences of words, which is a statistical model and also the basis for many natural language processing tasks [42]. Almost all LLMs based on neural networks use very large model architectures, such as 175 billion parameters [43], and are trained on large data sets, such as 1 TB English text [44]. This scaling enhances the ability of LLMs to generate coherent natural language [45-46], and allows them to be applied to a large number of other tasks [47-48] without modifying their parameters [43].

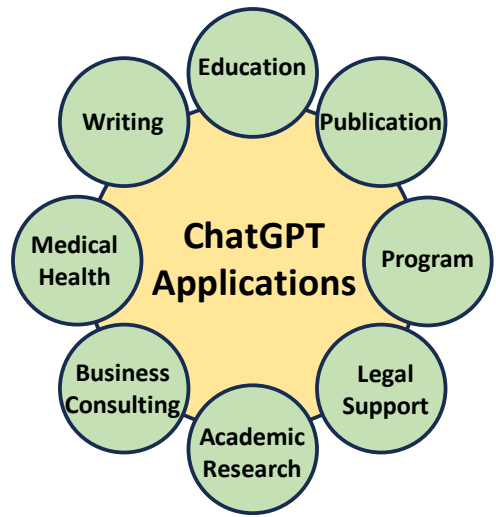

Fig 2. Typical applications of ChatGPT.

We illustrate the operation of a typical LLM ChatGPT using GPT-3.5 as an example. This model employs a stack of 13 Transformer blocks. Each block has 12 attention heads and 768 hidden units [15]. The inputs to the model are all a series of tokens that are first embedded in a continuous vector space by the embedding layer. It then inputs the embedded into the first block, which applies self-attention and generates hidden representation sequences [49]. GPT-3.5 further passes the hidden representation to the remaining 12 Transformer blocks. Each block includes a self-focused layer and a feedforward layer. The final output block is a hidden representation sequence that is decoded into an output sequence [50]. GPT-3.5 includes several add-ons in addition to the core Transformer architecture, such as layer normalization, residual connectivity, positional embedding, and so on. These components all contribute to stable training and improve model performance on language tasks. As summarized in [3], ChatGPT has been widely used in many scenarios, as shown in Fig. 2.

Based on GPT-3.5, OpenAI has released GPT-4. It is a multimodal LLM that can accept images and texts and generate outputs [51]. Meanwhile, it has demonstrated human-level performance on a variety of professional topics [52]. For example, it scored in the top 10% or so of a mock bar exam, better than GPT-3.5, which scored in the bottom 10% or so. In summary, GPT, as a representative of the LLM, has achieved remarkable performance on a large number of NLP tasks, including text generation, paragraph translation, and so on.

2.3 Blockchain for AI

Before entering the BC4LLM, we need to first understand the AI technology enabled by blockchain.

Many of the bottlenecks of AI can be effectively overcome by combining the two technologies [53-54]. AI relies on data to learn, infer, and make decisions. AI works better and more accurately when it learns from a secure, trusted data repository. Meanwhile, as a distributed ledger, blockchain data can be transacted and stored in a way that is cryptographically signed, verified, and consented to by participating nodes. Therefore, blockchain data has high security and integrity. Then, blockchain can create a highly secure, trusted, decentralized system for highly sensitive information in AI. As summarized in [27], blockchain-enabled AI systems will have many advantages, including data security, trusted decision-making, distributed intelligence, high efficiency, etc. In this way, distributed security intelligence is about to become possible.

Currently, the hottest area for applying blockchain is FL [28]. FL [55] was first proposed for training data distributed across multiple clients, and has been applied to data sharing, speech and image signal processing, etc. Traditional FL is a master-slave architecture, which uses a single master to aggregate data from a large number of distributed nodes. Therefore, the master node is easy to become the weak point of the system, and there is the possibility of a single point of failure. The decentralization and security of blockchain make it an important complement to FL, which further develops decentralized training models [56]. Zhou et al. [57] have proposed a blockchain-based distributed ML called Byzantine-resilience to protect the security of the training model. In addition, Jin et al. [58] have used cross-chain technology to enable FL and applied it to cross-clustering methods to improve aggregation efficiency. In [59], the authors have combined sharding technology and FL to propose a new computational method ChainsFL, which has high security. In [60], the authors have used blockchain, FL, and cryptographic primitives to propose a secure PPFchain model and apply it to wireless sensor networks.

At present, blockchain-based AI or FL has been applied to IoV [61-62], identity authentication [63], privacy computing [64], digital twins [65], and other fields due to its high efficiency and security. Thus, they also give us examples and motivation to design high-security LLM with blockchain.

## 3. Why do large language models need blockchain？

In this section, we will explain why LLM needs blockchain from three aspects: learning corpus, training process, and generating content.

### 3.1 Learning corpus

The realization of LLMs requires massive data and powerful computing power to support the training and reasoning process [66]. As the amount of data increases, the model can learn more features and patterns. For example, for image recognition tasks, more training samples can make the model learn more shapes, textures, colors, and other features; For natural language processing tasks, more training samples can enable the model to learn more features such as syntax, semantics, and context. Moreover, large-scale data can also reduce overfitting and improve the generalization ability of LLMs. For example, the ChatGPT has 175 million parameters, and the GPT-4 has 100 trillion parameters [15]. Since the learning corpus of LLMs is very large and important, it is necessary to find out the source of the data before analyzing the learning corpus of LLM.

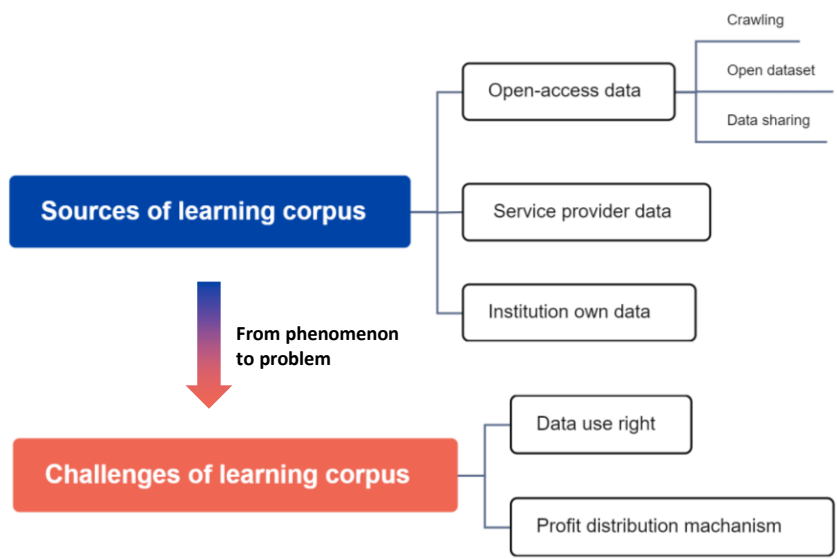

Fig 3. Sources and challenges of learning corpus.

### 3.1.1 Sources of learning corpus

At this stage, corpus sources of LLMs are mainly divided into three types, including open-access data, service provider data, and institution-own data [67], which are shown in Fig. 3.

1) **Open-access data**

After decades of information accumulation, the existence of a large number of government, institutional, and enterprise open-access data on the Internet can provide sufficient "materials" for LLM training. The crawler, open dataset, and data sharing are the main ways to obtain open data on the Internet.

a) Crawling refers to analyzing the structure of the web page and using the corresponding crawler to parse HTML, CSS, JavaScript, and other web content to obtain the required data. This type of data often requires further combing and labeling by professionals and is less available.

b) Open datasets are data publicly released by relevant organizations and institutions for use by researchers and developers, often including multiple domains and types of data, such as images, text, audio, etc. Common open datasets include ImageNet, COCO, OpenAI GPT, etc. This kind of data set is usually filtered, cleaned, labeled, and pre-processed by professionals, which has high quality and availability.

c) Data sharing refers to the exchange of data for mutual benefit. Unlike open datasets, data sharing usually provides more functionality and is often obtained in cooperation with data providers, such as Kaggle and UCI Machine Learning Repository. Such platforms allow users to upload, share, and access data sets, and provide tools and resources to facilitate research and development. This type of data is more professional and available than the above two.

2) **Service provider data**

Data service providers usually have a large number of data resources and technical strength, as well as can provide high-quality and diversified data services. They usually undergo professional screening, cleaning, labeling, and preprocessing operations to ensure the quality and availability of data sets. Meanwhile, they quickly deliver data sets on demand, reducing the time and cost of data acquisition for LLMs while easing the workload of data processing. In addition, the data service provider will also provide compliant and secure data services to avoid violating relevant laws, regulations, and ethical guidelines.

By working with data service providers, LLMs can obtain data sets that meet their needs and standards, avoiding the complexity and uncertainty of processing data. In other words, such data service providers can provide customized data services through LLMs.

3) **Institution-own data**

Own data refers to data sets that an organization or business owns that it collects, accumulates, and maintains. Enterprises use their own software to describe user portraits in a variety of ways to achieve accurate collection of user data. This type of data is very goal-oriented and focuses on the deep collection of a certain attribute or characteristic. In our lives, almost all Internet software requires users to open data rights, otherwise, it cannot use all functions. Every operational behavior of the user and data input may be used for LLM learning to further optimize its own parameters. The recommendation algorithm is a typical example, which analyzes every behavior and feature of the user to achieve accurate information feeding [68].

### 3.1.2 Challenges of learning corpus

With the cautious attitude of governments towards LLMs such as ChatGPT and the introduction of various data management regulations, data security has received more and more attention. Users are also gradually realizing the huge potential and value of data. It brings two aspects of data security problems, including data use rights and profit distribution mechanisms, which are also shown in Fig. 3.

1) **Data use right**

It is difficult for users to control the specific data use rights in LLMs. Specifically, in the traditional Internet operating model, data is mainly stored in the Internet company's cloud. Before users use App services, companies will force users to agree to privacy treaties, making it difficult for users to master data ownership. In addition, users lack effective means to track data. All user private data is in the hands of Internet companies, no matter how much they say they will not divulge data, but it is also difficult to monitor the scope and purpose of data used by companies.

At present, more and more intelligent services are dependent on Internet companies using user private data for machine learning analysis. Unfortunately, in this process, users are passive and have the suspicion of being violated.

2) **Profit distribution mechanism**

LLM, as a production tool, brings productivity improvement and production relations change to human society. While lowering the threshold of computer interaction, it also improves the efficiency of repetitive production for humans. The resulting huge productivity gains need to be redistributed between the big model owners and the data owners.

LLMs are largely centralized, and usually controlled by a few large companies or organizations. The user private data, as an important subject of the LLM learning corpus, deals with absolute disadvantages in the distribution of rights and interests. In addition, users cannot decide the scope and purpose of data use, so it is difficult to obtain the rights and interests they deserve.

**3.2 Training process**

The training process is the key to getting an AI system to learn how to perform a specific task. Trained with large amounts of data and algorithms, AI can continuously optimize and improve itself, thereby demonstrating superior performance in various fields. Before analyzing the security problems existing in the training process of the LLM, it is necessary to investigate the current training modes.

**3.2.1 Training mode status of LLM**

At present, centralized architecture is still the mainstream of traditional AI. Its objectives, technologies, systems, and services have the characteristics of concentration and static. When a system hosts AI tasks, a centralized architecture requires a central controller, regulator, coordinator, or communicator to manage and execute task operations. This will lead to basic problems such as system inflexibility, high vulnerability, poor reliability, scalability, and adaptability when centralized AI faces complex requirements and large-scale problems [69]. Furthermore, centralized AI also faces the problem of over-reliance on vendors for centralized aggregation of data in a single data center.

As a result, training a large and capable LLM using traditional centralized architectures is really challenging. In other words, for learning the LLM parameters, a distributed training algorithm is required, in which multiple parallel strategies are often combined [70]. To support efficient and accurate distributed training, some optimization methods have been published to facilitate the deployment of parallel algorithms, such as DeepSpeed [71] and Megatron-LM [72-73].

Recently, FL or distributed ML is the mainstream distributed training solution, which is expected to accelerate the training efficiency of LLMs [74-76]. Its training model is often similar to Fig. 4. Thus, to better use blockchain to solve the security problems in the LLM training process, we need to investigate the data security problems existing in FL or distributed ML.

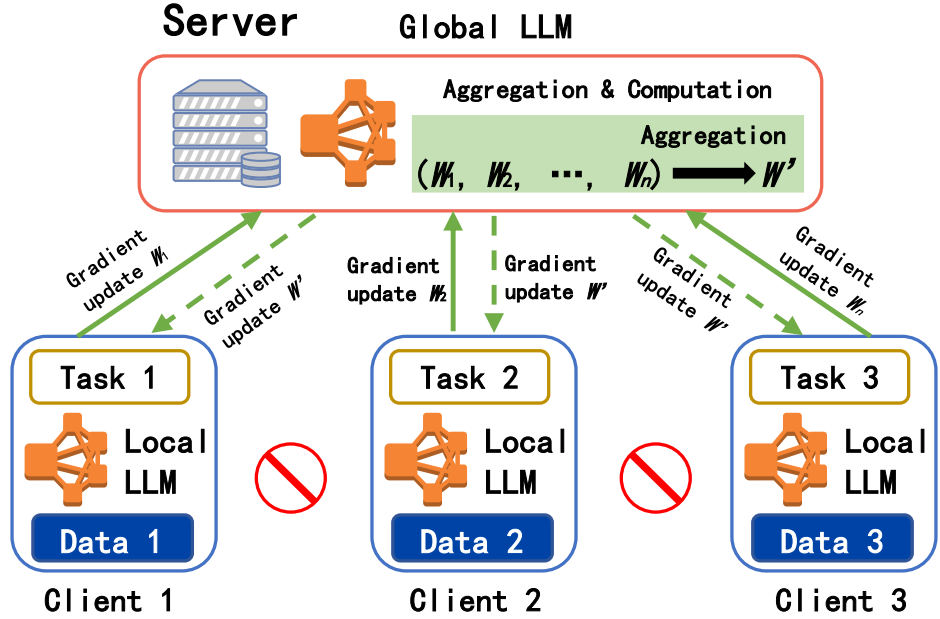

Fig 4. Distributed training for LLM.

#### 3.2.2 Challenges of training process

Inspired by [75], we divide the challenges of the training process into security threats and data privacy.

**1) Security threats**

Mothukuri et al. [77] gave a comprehensive overview of FL security threats, that is, attackers may exploit vulnerabilities and cause system security damage. Common attack methods include the poisoning attack [78] and adversarial sample attack [79].

a) Poisoning attacks can be further divided into data poisoning and model poisoning. Data poisoning means that the LLM learning corpus is incorrect and unsafe, resulting in an impaired training effect. This step occurs before training. In contrast, model poisoning is when an attacker injects malicious parameters or gradients into the global model to mislead the training process, thereby affecting the collaboration between clients. Due to the deep Transformer architecture and multi-stage training process of the LLM model, poisoning attacks have a high success rate and are more difficult to detect in distributed LLM training.

b) Adversarial sample attacks are mainly carried out in the inference stage. This attack often deliberately inputs data with subtle perturbations to deceive a trained model and lead to incorrect predictions. Due to the distributed nature of FL or distributed ML, global model parameters are shared in horizontal scenarios, which increases the possibility of model parameter leakage. This attack can be implemented after a distributed LLM deployment.

**2) Privacy disclosure**

Privacy disclosure refers to the harm caused by unauthorized access of external users to sensitive information such as local data and model parameters. Privacy attacks of distributed LLM may include several aspects, including learning corpus-based attacks, inference-based attacks, and prompt-based attacks, which are shown in Fig. 5.

a) Learning corpus-based attacks mean that training data from an LLM inadvertently appear in generated content, which may contain user private data, such as medical records and transaction amounts.

b) Inference-based attacks refer to the memory property of LLMs that have made them vulnerable to privacy breaches in FL or distributed ML [80]. The reason is that the Transformer in LLM is more easily attacked by dishonest servers. The dishonest servers can extract user private data through Transformer defects [81].

c) Prompt-based attacks are those in which certain promoting methods can induce LLM to output user private data. Attackers can even obtain more model permissions and access to user private data through special prompts. Therefore, this attack strategy can cause serious privacy disclosure.

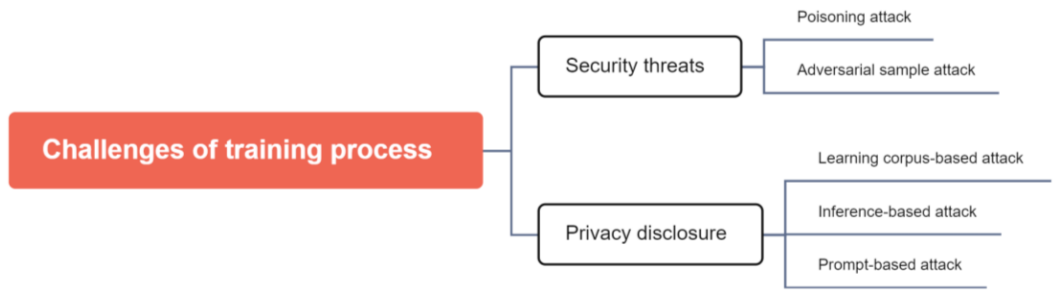

Fig. 5. Challenges of the training process.

## 3.3 Generated content

AIGC refers to a data production mode based on NLP, deep learning, generative adversarial network, and other AI algorithms to learn and simulate a large amount of data, so as to realize the understanding and mastery of natural language rules, and automatically generate text, pictures, speech, and other content [82-83]. It can achieve innovative content generation in a cost-effective and efficient way according to the customized needs of users [84-85].

These novel AI-generated content, some of which are "patchwork creations", seem to be different, but are actually related to existing Internet content; Some are "created out of thin air", and completely subvert our cognition and creativity. Being able to uniquely identify AI-generated content and query its creators and owners would provide a better solution to the credibility and regulatory compliance of AIGC [86-88]. As a consequence, we need to investigate the identification scheme for AI-generated content.

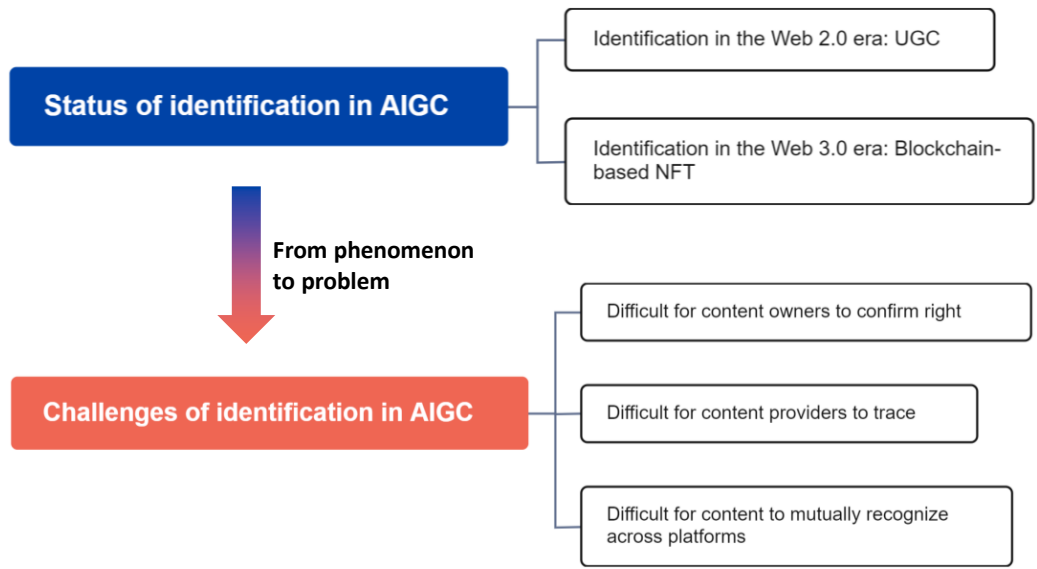

Fig. 6. Status and challenges of identification in AIGC.

**3.3.1 Status of identification in AIGC**

We have divided the status of AI-generated content identification into the Web 2.0 era and Web 3.0 era to investigate separately, as shown in Fig. 6.

1)  **Identification in the Web 2.0 era**

User Generated Content (UGC) is an online information resource creation model born in Web 2.0 [89]. It means that users display their original content through the Internet platform or provide it to other users. UGC has three basic characteristics, that is, based on the premise of network publishing; The content has a certain innovation; And the content is created by non-professionals or non-authoritative organizations.

With the development of AI technology, user-generated UGC content is also gradually moving to user-generated AIGC content. For example, the virtual user avatar in the short video of TikTok, the cartoon avatar generated by the user actual photo in WeChat, etc. The identity of user-generated AIGC content is usually generated by the Web 2.0 platform. The source file of the content is stored in the platform server and transmitted by a URL form, which is convenient for users to download and upload. This kind of identification pays more attention to information production's individuation, immediacy, and interactivity.

2)  **Identification in the Web 3.0 era**

Since entering the Web 3.0 era, we have often used blockchain to identify AI-generated content [90]. A Non-Fungible Token (NFT) is a typical example. It is a unique set of digital identifiers that cannot be copied, replaced, and subdivided based on the blockchain smart contract and can be used to prove the authenticity and ownership of the token content [91].

CryptoPunk [92] is considered to be the first combination of AIGC and NFT. Each of the cryptopunk avatars is generated according to different pixel combinations, and all use NFT as content identification to achieve ownership verification. Then, the Dapper Labs officially launched the ERC721 standard specifically for NFT [93], and launched the Crypto-Kitties game based on it. Each encrypted cat picture is generated by genetic information through AIGC technology, and the corresponding NFT records the genetic information, reproduction algebra, and ownership information of each encrypted cat picture. Nowadays, more blockchain-based identification schemes have been proposed, such as ESIA [21], BIDaaS [94], NIDBC [95], and so on.

**3.3.2 Challenges of identification in AIGC**

At this stage, the identification technology of AI-generated content has the following three challenges, that is, difficult for content owners to determine rights, difficult for content providers to trace, and difficult for content to mutually recognize across platforms [96], which are also shown in Fig. 6.

**1) Difficult for content owners to confirm rights**

AI-generated content is represented as data that can easily be stolen and copied. Even if the AIGC identity with blockchain participation has on-chain confirmation ability, and can prove the content ownership through the NFT, users can still copy and download the corresponding content at will according to its link. The reason is that NFT is still mostly presented in the form of URL identification on mobile and web sides. This problem is more common in the UGC scenario without blockchain participation, especially for content ownership determination [97]. For example, in 2018, financial articles created by Dreamwriter, an intelligent writing assistant independently developed by Tencent, were forwarded by websites operated by others. After that, Tencent sued for unauthorized reprinting, wasting a lot of manpower and time.

**2) Difficult for content providers to trace**

Identity of AIGC is typically used only to annotate the generated content, by embedding the identity as part of the data in the generated content. Content providers can control the generation and use of this data and determine how data is transmitted and stored. However, due to the manual orientation of the AIGC training model, it is easily used by malicious content providers to develop a variety of products and services with more extreme ideologies. If these products and services are used by special user groups, they will have a negative social impact. In practice, ownership of AI-generated content is usually determined by the user rather than the provider [98-100]. In this case, if there is a violation of AI-generated content, the responsible person can only be traced back to the content user and not to the content provider. Consequently, the "initiator" of malicious AI-generated content cannot be accurately traced, which will bring many adverse factors to the health and stability of the Internet ecology.

3) Difficult for content to mutually recognize across platforms

AIGC has used a variety of technical methods and carrier means of content identification, including digital watermarking [101], content-embedded hash value [102], and so on. For example, digital artist Trevor Jones uses digital watermarking technology in his work to ensure copyright and provenance However, different content identification technologies have their own advantages and disadvantages. instance, digital watermarking can embed identification information without affecting content quality, but it is easy to be attacked; Hash values guarantee uniqueness and completeness, but they do not provide more information. In addition, different identification technologies use different coding schemes, such as NFT class AIGC using longer hash identifiers, and UGC-oriented identifiers usually using the form of random generated by platforms or user self-naming. Therefore, with the rapid development of AIGC applications recently, due to the lack of unified standards for generating content identification, resulting in the interoperability and interoperability between different content identification technologies need to be solved.

4. How can blockchain enable large language models?

According to the contents sorted out and the problems to be solved in Section 3, this section will also discuss how blockchain enables LLMs from three aspects: learning corpus, training process, and generated content. In other words, we have enabled a reliable learning corpus, a secure training process, and identifiable generated content via blockchain to build trusted AI LLM. The overall architecture is shown in Fig. 7.

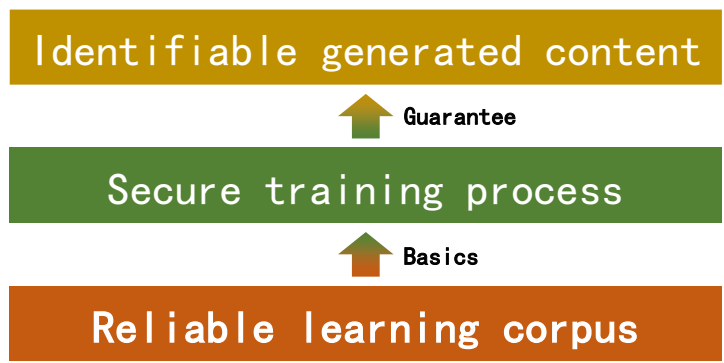

Fig. 7. Technical route for BC4LLM.

## 4.1 Reliable learning corpus

When users experience AI services, they also face the risk of data leakage and abuse. Blockchain, as a decentralized and trusted enabling technology, can effectively realize the data confirmation and privacy protection of LLMs. We discuss the function of blockchain for LLMs from two aspects: data right confirmation, and restructuring profit distribution.

### 4.1.1 Data right confirmation

In the AI and big data era, the ownership of digital assets has become an important concern issue. Confirming ownership of digital assets is a necessary prerequisite for encouraging data owners to contribute their data, which also provides LLMs with more available and valid data, learning and training materials.

The traditional method of data right confirmation adopts the mode of submitting ownership evidence and examining authority [103]. This approach lacks credibility due to poor supervision, and there are often uncontrollable factors such as potential tampering [104]. As a result, this issue is a long-term bottleneck for data sharing. Fortunately, the advent of blockchain has brought some new data-right confirmation models that can provide better data security.

A blockchain-based data authority confirmation model typically has three parts: a client, a blockchain network, and a blockchain ledger, as shown in Fig. 8. The client is responsible for the interaction with peers in the blockchain network. Each peer will be bound to a unique client. In addition, the client is initialized by the security component to ensure security. Security components tend to include four functions [103]: generating public keys for users, extracting the data feature, providing an interface for chain code interaction, and submitting a data right transaction. Among them, the extracted data features depend on the design of the model. In [103], the authors used data fingerprints based on local sensitive hashes as data features, while in [104], the authors extracted digital features based on symbol mapping coding. Moreover, the blockchain network is used to transfer block and transaction data between peers to achieve consistency and keep data synchronized. The transmission mode is usually broadcast protocol, such as Gossip [105], Kadcast [106], and LECast [107]. Furthermore, the distributed ledger is jointly maintained by all peers of the blockchain. The benefit is to avoid data loss due to peer failure. All transactions are recorded on the blockchain in an immutable form, and these transactions interact through the underlying key-value database. Additionally, similar work includes BBS [108], which is designed for the secure sharing of very large data based on blockchain to protect user privacy and confirm data ownership.

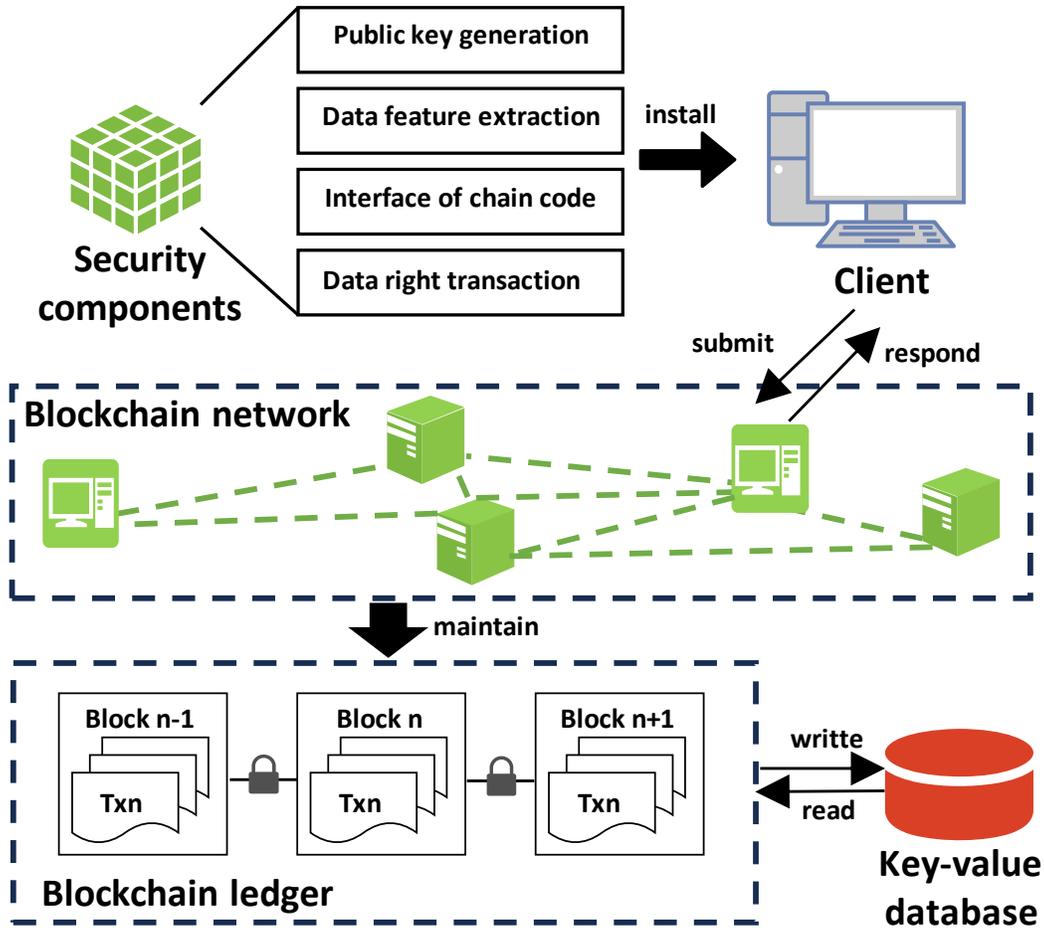

Fig. 8. Blockchain-based data right confirmation model.

**4.1.2 Restructuring profit distribution**

A reasonable profit distribution mechanism can encourage users to provide their private data for LLM development companies to learn and train. At present, the profits earned by LLMs are largely captured by their development companies and rarely shared with data providers, which severely limits the enthusiasm and motivation of data providers. It is important to note that profit also includes responsibility. Once the data rights are confirmed, if the data provided by the data provider is false, it will be traced and held accountable through the blockchain. That is to say, blockchain provides users the autonomy to choose according to the profit distribution perspective, thus, facilitating users to provide private data and ensuring data security. The collection, flow, storage, and sharing of these provided data are implemented on decentralized functional nodes in the blockchain to ensure the traceability and trust of training data. When LLMs obtain revenue, part of the revenue is allocated to the data provider according to the training data weight, so as to ensure its legitimate rights and interests.

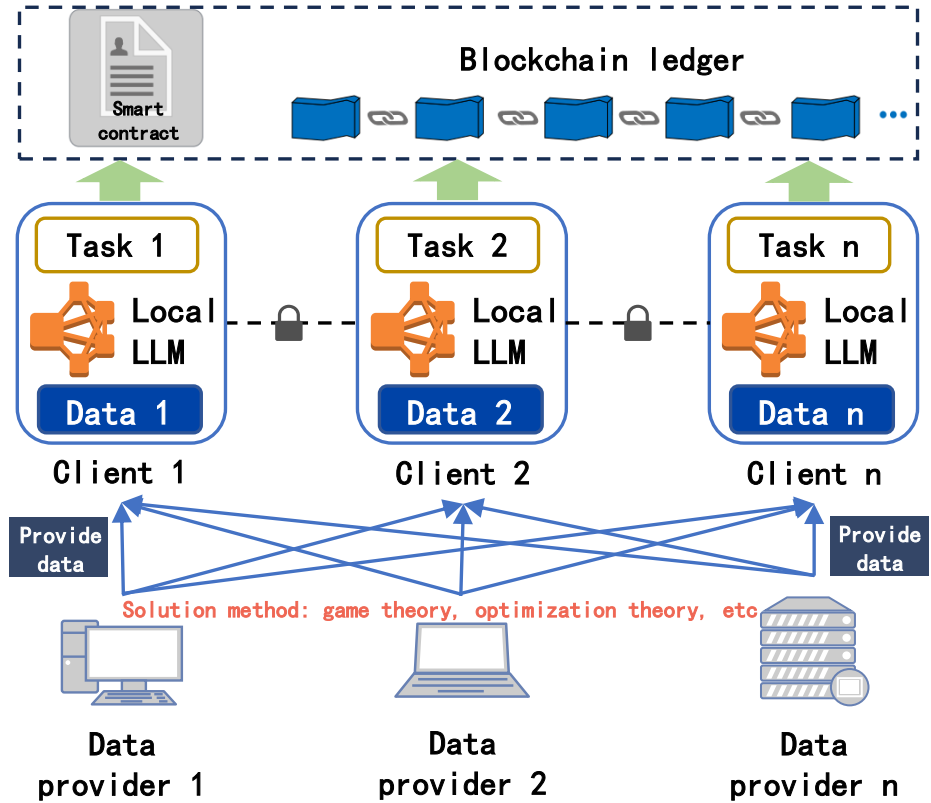

Fig. 9. Blockchain-based profit distribution architecture.

A representative work BABG [109] shows us a blockchain-based profit distribution mechanism in a Vehicle-to-Grid (V2G) scenario. All energy transactions for electric vehicles (EVs) are recorded on a blockchain ledger to ensure privacy. The scheme further uses the smart contract as a proxy for the energy and profit distribution. On this basis, the authors introduce a two-way auction mechanism based on the Bayesian game, and design a new price adjustment strategy. Similar work has also been included [110-111], where blockchain is combined with game theory to determine the best profit distribution. In addition, there are also studies on blockchain based on optimization theory to confirm the working utility of blockchain nodes [112-113]. The above scheme has been widely used in V2G, and wireless blockchain network scenarios, which also have the potential to be deployed in LLMs.

Fig. 9 shows our proposed blockchain-based profit distribution architecture for LLMs. Among them, a blockchain network is built through wired and wireless communication by clients undertaking different training tasks. For each client, data is collected, learned, and trained from multiple data providers. Each client will maintain a blockchain ledger to confirm the source of the learning corpus and the immutability of the training results. Such an architecture demonstrates the potential of blockchain to trace data in LLMs and can inspire more data providers to provide private data for LLM training.

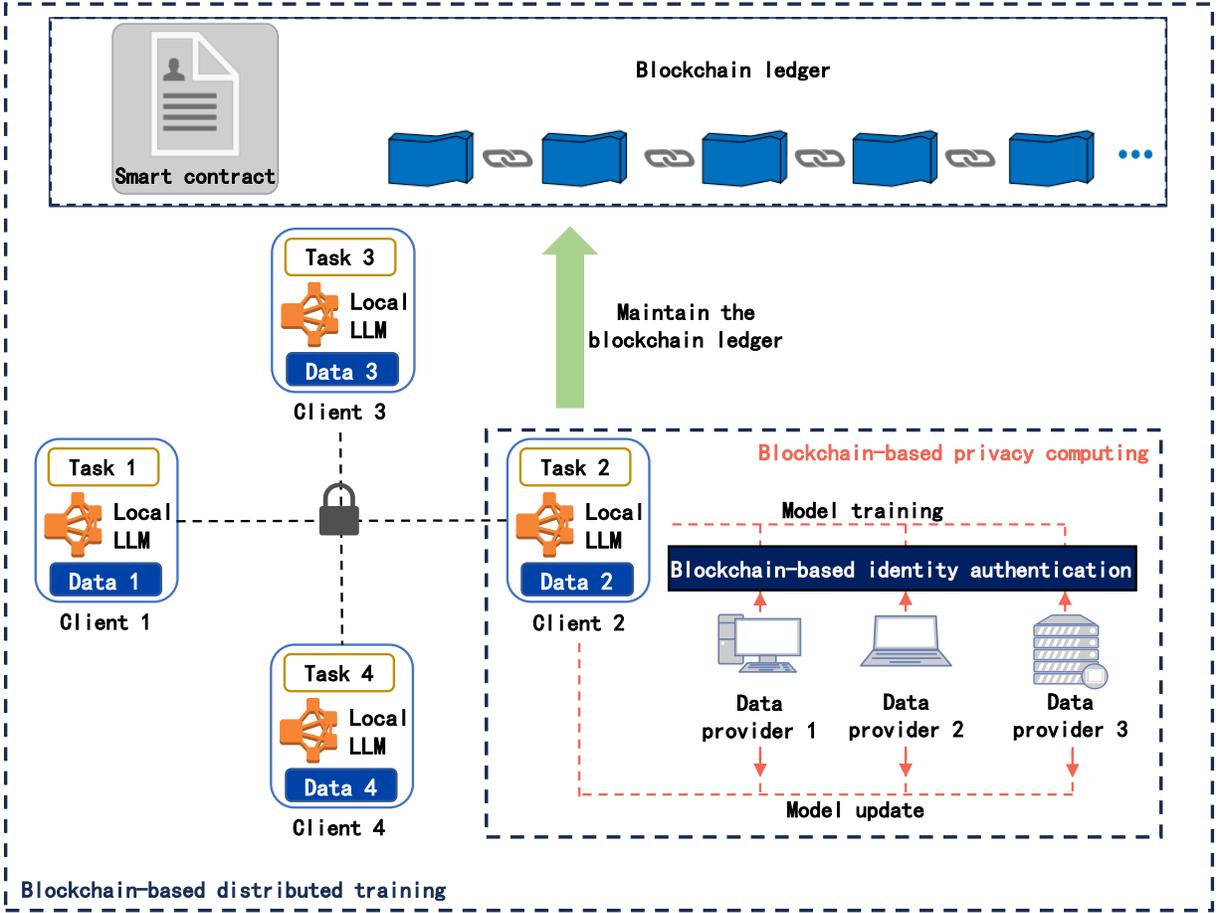

Fig. 10. Blockchain-enabled distributed training model.

**4.2 Secure training process**

According to the analysis in Section 3, to break the data silos between companies, institutions, and organizations and improve the training effect of LLMs, distributed training will become the mainstream paradigm of AI in the future [28, 63-64]. However, companies, institutions, and organizations do not want their core secrets to be held by others, so we need to use blockchain to design secure and trusted distributed training methods to address privacy breaches in distributed LLMs [65, 114].

**4.2.1 Secure distributed training**

The most important thing for a secure training process is a secure training model. Such models often build each client as a distributed blockchain network and maintain the same blockchain ledger to ensure the consistency and immutability of training results [28], which is shown in Fig. 10.

In [115], the authors combine various cryptographic methods, such as differential privacy and homomorphic encryption, to design a blockchain-based trusted collaborative distributed learning scheme. In their design, key steps of collaborative distributed learning are recorded on the blockchain, which can effectively track and block malicious behavior in a timely manner. In addition to protecting private data, blockchain-enabled distributed ML can also resist a variety of attacks, such as Byzantine attacks. Li et al. [28] have proposed an anti-Byzantine consensus, named Proof-of-Accuracy (PoA) to implement a distributed ML that can detect and resist Byzantine attacks. Meanwhile, the authors have improved the efficiency of model verification by introducing validators to perform heavy verification work. Furthermore, blockchain-based distributed training models can also avoid malicious propagation in the network and can be applied in scenarios where the credibility of participants is unclear [116]. This model can realize the security protection of mobile distributed machine learning systems. There are many other models similar to the above scheme, such as [60, 117-118], etc.

**4.2.2** Identity authentication and privacy computing

The above work is to build each client as a blockchain network for distributed training, to achieve the purpose of secure training for LLM. In addition to that, different technical routes also include blockchain-enabled FL or distributed ML identity authentication [63], privacy computing [64], etc, as also shown in Fig. 10.

Secure identity authentication can endorse the identity of the client and data provider participating in the training, further enhancing the security of the training process and results. In [63], the authors have combined directed acyclic graph (DAG) blockchain and accumulator to design a decentralized and simplified FL authentication framework to achieve efficient and secure identity authentication. In [119], the authors have utilized a flexible blockchain consensus and zero-knowledge proof to verify the identity of every participant in FL, which can achieve high training accuracy. Additionally, Chatterjee et al. [120] have considered the need to design a strict blockchain-based identity authentication mechanism for FL participants in the financial transaction scenario to ensure financial security.

Privacy computing ensures the security of each client during training. Meanwhile, the client submits the block to the chain in a confidential form and updates the data and parameters in a secure form. In [64], to achieve lightweight privacy protection, the authors have adopted Paillier encryption and designed a lightweight digital signature to make privacy computing more efficient. Meanwhile, in [121], the authors have used a secure global aggregation algorithm in the blockchain-based FL architecture to defend against malicious devices and avoid privacy breaches during computation. Moreover, Bhatia et al. [122] have provided a way to use blockchain to ensure the quality of FL training model data, which can guarantee the accuracy of training results and ensure that privacy is not compromised.

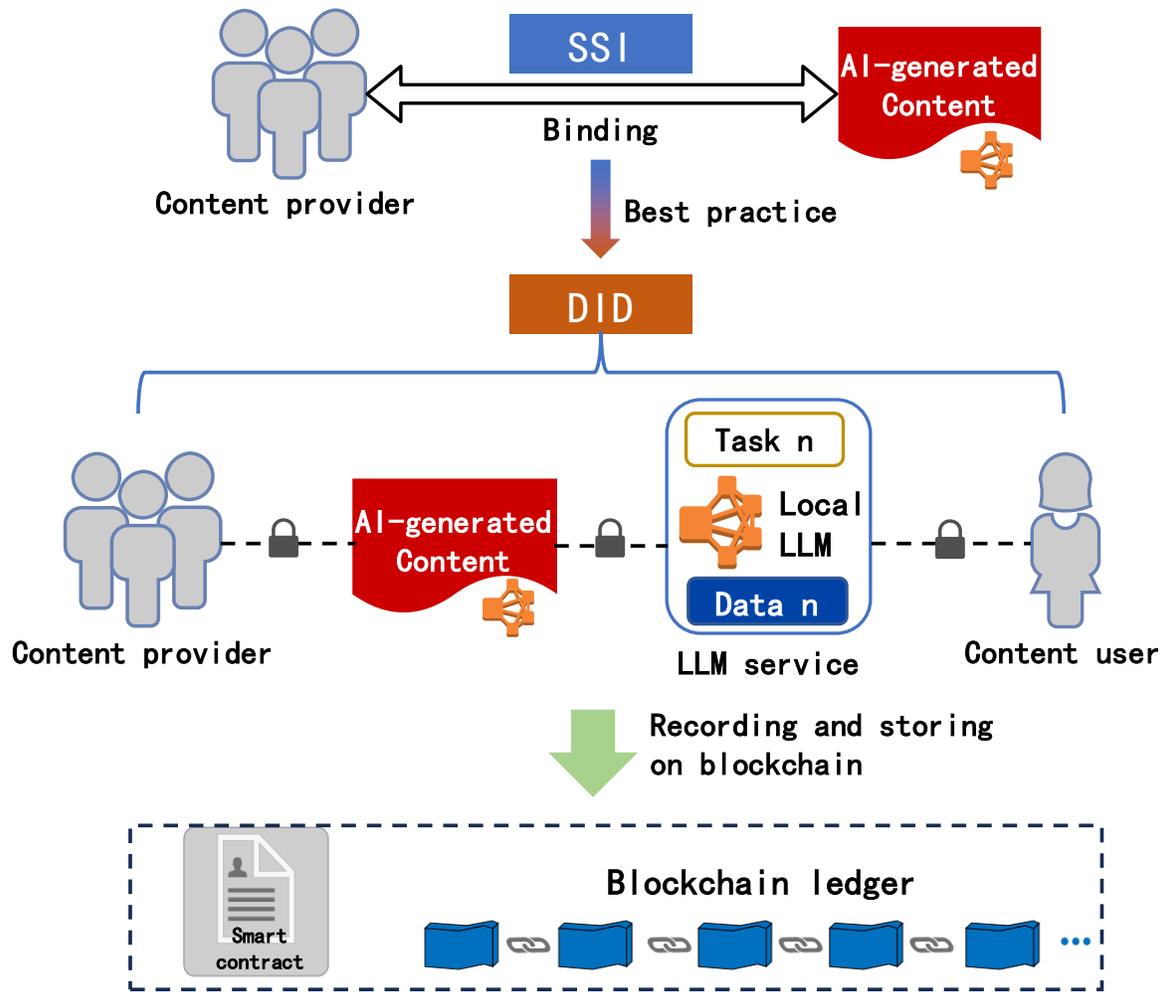

Fig. 11. Blockchain-enabled identification service.

**4.3 Identifiable generated content**

Identifying and confirming AI-generated content is a way to respect the content provider's work and trace the content circulation. This can also help the content provider endorse the quality and authenticity of the generated content and help the circulation of the AI-generated content. Therefore, we use the blockchain to identify and confirm the right to the AI-generated content, and bind the generated content and its provider. It should be noted that the confirmation right to AI-generated content here is not the same as the confirmation right to data in Section 4.1.1. Here is the binding to the generated content and its owner[1], while in Section 4.1.1 is the binding to the LLMs learning corpus and its owner.

---

[1] Here, we treat the owner and provider of content as the same role.

The technical route is divided into three steps, first using Self-Sovereign Identity (SSI) [123] to generate the identity of AI-generated content, then, relying on the Decentralized Identifier (DID) [124] to establish a unified identity service system, as shown in Fig. 11.

**4.3.1 Self-sovereign identity**

A secure and reliable identity is necessary to accurately identify users. SSI is an emerging and important means of achieving this goal. It is realized possible by the rise of blockchain technology, which replaces the classic identity management registration process through a trustless decentralized database provided by the blockchain [123]. Since the autonomy of its management, it has been widely used in data-sensitive scenarios such as healthcare [125], data management [126], and sustainable development [127]. In [128], the authors have further pointed out seven functional requirements that SSI should meet, namely (1) Being able to communicate directly with other peers; (2) Authentic; (3) integrity; (4) network level anonymity; (5) Decentrally synchronize the information published by the directory server; (6) Identify with oneself in a self-sovereignty way; (7) Accountability to the subject.

Once SSI achieves this vision, then the identity for AI-generated content we built on SSI can effectively realize the confirmation right to the content provider and the traceability of content circulation. Since the AI-generated content usually consists of the user individual needs and the LLMs generation service, the identification of the generated content can more flexibly and conveniently build the trust relationship between users, LLMs, content providers, and other parties, and meet the regulatory compliance requirements.

In addition, identifiers for AI-generated content built on SSI can return content ownership to the user. This allows for a more flexible approach to AI-generated content management. Users can generate, assign, and manage their own content identifiers to ensure the uniqueness and trustworthiness of their content, while reducing the possibility of third-party intervention or management.

**4.3.2 Decentralized identifier**

DID is a specific technical solution that is promoted by W3C, the World Wide Web Consortium of the international standardization organization, and practices the concept of SSI. The identifiers of AI-generated content based on DID can provide globally unique identifiers and build a unified identifier service system based on the existing DID identification parsing technology [129]. Since DID is now the best de facto standard for implementing SSI concepts, doing so allows for both better interoperability and portability [124]. We can more accurately construct LLM participants and their interactions by assigning specific DID identifiers to each AI-generated content and corresponding content providers. In addition, we can also achieve cross-platform and cross-application interactive sharing of AI-generated content more conveniently through standardized DID identity parsing protocols. To implement DID, the authors [124] have stated that it should contain three important functional components, including issuer, receiver, and verifier.

Applying blockchain to the identity design for AI-generated content can provide it with a verifiable, immutable, decentralized identity key infrastructure [130]. Specifically, blockchain can be used to record the identity and key information of AI-generated content, truly ensuring authenticity and accuracy from the source of the data. For example, the identity of each AI-generated content is realized by DID, and then the identity, content provider, and content user are recorded in the DID document and stored on the blockchain, which can form a decentralized and immutable on-chain record. In this way, it ensures the authenticity of AI-generated content, content providers and users, and LLM services. At present, due to the above benefits of DID on identity, it is applied to energy trading [131], source address verification [132], etc., showing great potential in LLMs.

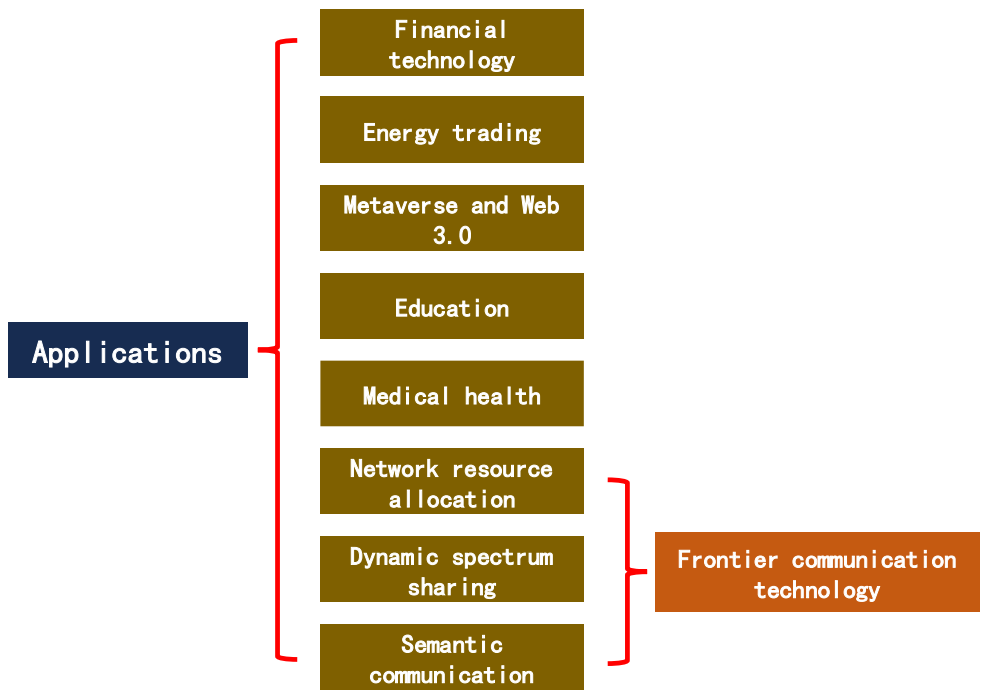

Fig. 12. BC4LLM applications.

## 5. Applications of BC4LLM

Recently, the LLM represented by ChatGPT has been widely used in many fields. As an LLM combined with blockchain, BC4LLM has more secure and trusted attributes, which will enable more data-sensitive scenarios, such as the financial sector. Moreover, blockchain can also guarantee the authenticity and accuracy of LLM training results [133], so that BC4LLM can also be applied to scenarios with high demand for the authenticity of prediction results, such as education and medical health. The related application scenarios are shown in Fig. 12.

## 5.1 Financial technology

At present, the AI models used in the financial industry are mainly "small models", which rely on expert experience. The amount of training data and the number of parameters is small, and its advantage is "specialized" and can meet the specific task requirements.

However, with the continuous development of technology, LLM will also play an important role in the financial field in the future. For example, the LLM can mine and analyze large amounts of financial data to help financial institutions more accurately assess credit risk, identify fraud, and prevent financial crimes. Meanwhile, speech recognition and natural language processing technologies in LLMs can improve customer service quality and business processing efficiency. In addition, LLMs can also provide investors with more accurate and personalized investment advice through data analysis and forecasting to help investors maximize their assets. Nowadays, there are relevant research efforts to apply LLMs to fintech, such as sentiment analysis [134], news classification [135], and question answering [136].

There is no doubt that the financial sector is most concerned about data security and privacy protection of customers and financial transactions, which is the cornerstone of providing great financial services [137]. Therefore, BC4LLM with blockchain capabilities will further provide security services for fintech on the basis of LLMs. BC4LLM enables complete confidentiality of customer data and immutability of financial transactions.

## 5.2 Energy trading

The combination of AI and blockchain is relatively new in the energy sector, with only some early work on AI [138-139].

BC4LLM can first play the role of AI in the field of energy trading. It can conduct deep learning through historical energy price data and other relevant factors (such as supply and demand, global economic situation, etc.) to predict future energy price trends, helping energy traders better grasp market dynamics and improve trading returns. Additionally, BC4LLM can analyze historical energy trading data and market changes, predict future risks, and provide traders with corresponding risk management recommendations and measures. Furthermore, BC4LLM can also help energy reserve institutions better forecast reserve demand, avoid losses caused by energy price fluctuations, and improve energy reserve efficiency.

In addition to the AI function, BC4LLM can also play the blockchain role in energy trading. On the one hand, energy trading, like finance, is an area that is extremely sensitive to trading data. The reason is that the electricity consumption data on the user side can be used to monitor the living conditions of the house, which may be used by thieves; On the power generation side, if the power generation data is stolen by competitors, it will bring vicious business behavior [140]. Therefore, the energy sector also needs blockchain to thoroughly protect its private data. On the other hand, smart contracts in blockchain can help traders better manage energy trading, including automatic execution, monitoring and adjusting contract terms, etc., to achieve more efficient and accurate trading operations [141].

### 5.3 Metaverse and Web 3.0

Both blockchain and AI have been regarded as the native technologies of the Metaverse and Web 3.0, and will contribute to their intelligence level and operational efficiency [142]. BC4LLM will further strengthen this function.

The LLM will enable all Metaverse and Web 3.0 systems, and provide them with technical support beyond the learning level of humans [143]. For example, the voice service in LLMs will provide users in the virtual world with accurate voice recognition, and communication technology. In addition, LLMs also provide more accurate prediction and data support for the virtual world construction in the Metaverse and Web 3.0, as well as the interaction between the virtual world and the physical world, which can significantly improve the user experience [144].

Meanwhile, blockchain, as one of the infrastructures in the Metaverse and Web 3.0, can connect isolated small sectors together to provide a stable virtual world and reliable and secure rules. For instance, hash algorithms and timestamps in the blockchain can make metadata traceable and confidential [145]. Moreover, digital currency and NFT transactions in the virtual world will also involve financial security, and it is urgent for blockchain to play a role in protecting transaction privacy. In addition to the virtual assets security, the distributed storage of blockchain can also guarantee the user's identity in the virtual world [143].

### 5.4 Education

LLM has been widely used in the education [8-10, 16]. On the one hand, chatbots represented by ChatGPT can assist students in generating writing tasks; On the other hand, LLMs can also help teachers and professors to accurately analyze students' performance and scores, so as to teach students according to their aptitude.

However, the first problem with this is that the content generated by AI may come from false information, resulting in inaccurate or even wrong knowledge for students. Second, students' over-reliance on AI-generated content will also bring serious plagiarism risks and affect their academic integrity. The third problem is the risk of data breaches when teachers and professors use LLMs to analyze student scores. Once the score data is obtained by external institutions, it will cause distress to students, such as targeted advertising and push.

BC4LLM with blockchain sewn will effectively solve the above problems. BC4LLM first identifies data sources and generates reliable reference content for students through trusted data. Second, blockchain-based content validation and confirmation can also identify and trace AI-generated content to avoid plagiarism by students. Finally, a blockchain-based grade management system will effectively eliminate the possibility of score data leakage.

**5.5 Medical health**

Ref. [11, 58, 97] has given us a broad picture of how AI and blockchain can be applied to healthcare. We can use LLMs for diagnostic assistance to help doctors improve the accuracy and efficiency of disease diagnosis. By analyzing medical images, LLMs can automatically identify abnormal areas and provide auxiliary interpretation to reduce missed diagnoses and misdiagnoses. In addition, LLMs can be used to predict and evaluate diseases, helping doctors to develop more accurate treatment plans. By analyzing a patient's clinical data and biomarkers, LLMs can predict the development trend and disease risk, and provide doctors with personalized prevention and treatment recommendations. Furthermore, in combination with sensor technology and smart devices [146], LLMs can also monitor patients' physiological parameters and activities, and promptly alert patients to adjust treatment plans or seek professional medical help.

However, as a data-sensitive area, medical data is extremely private for every patient and cannot be accessed outside the medical establishment [147]. Fortunately, the blockchain capabilities provided by BC4LLM eliminate the above pitfalls. Blockchain-based access control protocols and identity authentication technologies provide a security line for access to medical data. In addition, distributed storage mode and data encryption of blockchain enable medical privacy integrity and secure storage. Meanwhile, the time stamp in the blockchain can also record the operation time and operator of medical data, playing the role of data traceability.

**5.6 Network resource allocation**

Network resource allocation is one of the key technologies in the communication networks field. As the number of devices connected to the network continues to increase, so does the data traffic. However, the available network resources are limited, thus, how to allocate and schedule network resources is the key to improve network service capability [148].

AI-based intelligent allocation of network resources has been applied in different network types, such as IoV [149]. With the rise of LLMs, the powerful computing power it provides will provide more efficient and accurate solutions for network resource allocation and ensure the quality of network service. LLMs can analyze network data to predict network congestion, faults, and anomalies, and take appropriate measures to optimize network performance in advance. These advantages are expected to promote the application of LLMs in more network types, such as wireless networks [150], satellite networks [151], drone swarms [31], and so on.

BC4LLM can build on the above and play the unique role of blockchain. It can provide a trading platform for various network resources, such as bandwidth, storage, computing power, etc. Through blockchain technology, the platform enables direct interaction between the supply side and the demand side of resources. In addition, blockchain can also be applied to the secure management of network resources. It not only provides a secure and trusted network environment to prevent network resources from being maliciously attacked or abused, but also encrypts and authorizes access to network resources. Finally, through the smart contract, BC4LLM can achieve automatic scheduling and allocation of network resources.

**5.7 Dynamic spectrum sharing**

Dynamic spectrum sharing is a necessary change in the field of wireless communication to cope with the growth of wireless services but limited spectrum resources [152]. This technology can make full use of the limited spectrum resources and avoid spectrum resource waste. In addition, it can also make different users share the same spectrum, thereby increasing the system capacity and improving communication efficiency. Therefore, it is also regarded as an enabling technology for B5G and 6G communication.

LLMs can play an important role in dynamic spectrum sharing [153]. In particular, through the analysis of network data, it predicts the spectrum demand of different networks when the business conditions change, and automatically adjusts the spectrum-sharing strategy to achieve more efficient spectrum utilization. In addition, LLMs can also be used to evaluate the performance of different networks under different spectrum conditions, such as transmission rate, delay, packet loss rate, and other indicators to provide a reference for spectrum sharing strategy.

The role of BC4LLM in dynamic spectrum sharing is reflected in many aspects [154-155]. First, blockchain can enable efficient management and coordination of the spectrum. As a distributed database, blockchain can securely record and update spectrum usage, avoiding issues such as single points of failure and data tampering. Second, blockchain can realize the trusted spectrum access mechanism through its decentralization and traceability system, ensuring the fair use of spectrum and the reasonable allocation of resources. Finally, complex cryptographic algorithms in the blockchain can be used to achieve the privacy protection of spectrum sharing, ensuring the security of user information and the compliance of spectrum use.

**5.8 Semantic communication**

The semantic communication is considered a break from the Shannon paradigm, which aims to successfully convey the semantic information of an information source, rather than receiving every symbol or bit accurately [156]. It has been identified as the core challenge of 6G communications [157].

However, in the existing semantic communication system, the knowledge base construction is faced with many problems, such as limited knowledge representation, and frequent knowledge updating [158]. In particular, in multimodal language communication, it needs to simultaneously transmit text, audio, and pictures, such pain points will be further amplified [159]. Fortunately, LLMs have evolved to provide new solutions to overcome the above problems. Specifically, LLMs can integrate semantic segments as new semantic awareness sources and build knowledge bases. In addition, LLMs utilize their powerful learning ability to perform tasks such as high-quality semantic segmentation, and can also be used for information extraction and reasoning in semantic communications [160].

Furthermore, the need for communication security promotes the rise of BC4LLM in semantic communications. The blockchain guarantees the security of the LLMs when extracting semantic information, avoiding interference by malicious attackers [161]. Second, blockchain is expected to make semantic communication systems decentralized to avoid a single-point failure in centralized semantic communication systems. In general, BC4LLM for semantic communication is still a new and urgent field.

**6. The future challenges**

As described above, BC4LLM has unique advantages in data quality verification, data right confirmation, content identification, decentralized AI model, and other aspects, which can be widely used in many fields. However, with the further development of LLMs, the scale of training parameters and data continues to grow, which will bring potential challenges and difficulties to BC4LLM.

Here, we summarize five important potential challenges: blockchain scalability, storage overhead, energy consumption, heterogeneous deployment scenarios, and model interpretability. Meanwhile, we also discuss the corresponding promising solutions.

**6.1 Blockchain scalability**

Blockchain consensus makes an important contribution to achieving consistency among the nodes in the network, and it is also the basis for consistent decisions and actions in a distributed network. However, the vast majority of consensus has multiple rounds of communication, which leads to a lot of communication overhead and delay, severely limiting the blockchain network scalability [162].

Meanwhile, LLMs often involve multiple parties, including a large number of data owners, content providers, and service providers. Then, the limited blockchain scalability will also limit the service BC4LLM scale. As a result, we must improve blockchain scalability to enable BC4LLM to access more participants, increasing the scale of learning and training data, and serving more users. Fortunately, technologies such as sharding, DAG chains, sub or side chains, and payment channel networks offer mitigation solutions for blockchain scalability. However, such methods should be further studied in the LLM field.

**6.2 Storage overhead**

Distributed databases make blockchain famous for avoiding single-point failure and achieving decentralization. However, such a mechanism makes every node in the blockchain system need to back up all the block and transaction data [163], which will lead to a lot of storage resource waste.

To make matters worse, LLMs have a huge amount of learning corpus, training data and parameters, and also generates a huge amount of AI content. These rich but redundant data will further increase the storage overhead of BC4LLM, and make the distributed database more burdensome. At present, there are related technologies to alleviate this problem, such as the hierarchical architecture for the blockchain, which can limit the storage overhead to the storage layer [164]. In addition, it is also feasible to reduce data redundancy through distributed coding [165]. However, blockchain with low storage overhead should be further studied for more complex applications such as the LLM.

**6.3 Energy consumption**

Both the consensus and storage of blockchain are functions that consume a lot of energy. In addition, the LLM requires large-scale training, which consumes more energy. Therefore, the energy consumption of BC4LLM is very prominent and extremely important.

Currently, there is a lot of work on designing energy-efficient consensus algorithms to achieve a sustainable blockchain [20, 166]. And there are also low-energy consumption broadcast protocols [107], and storage structures for the blockchain [167]. In addition, there are also some solutions for AI training, such as improving the learning framework [168] and using more powerful hardware [169]. However, this area is not well studied. In particular, the energy consumption of BC4LLM, which combines blockchain and LLM, is more worthy of in-depth discussion.

### 6.4 Heterogeneous deployment scenario

In a distributed training scenario, BC4LLM needs to connect various training clients through wired or wireless communication to build a blockchain network. The simultaneous existence of wired and wireless heterogeneous communication environments will have an impact on the transmission of training data and parameters, as well as block and transaction data in the blockchain network. In addition, the individual training clients are not always online and may be offline, which can also cause confusion in the information transmission of the blockchain network.

As a consequence, heterogeneous blockchain networks will adversely affect training efficiency and results. At present, there is not much research on such issues. There are only a few articles on wireless blockchain network performance analysis and modeling [22, 150], as well as work on designing blockchain consensus in wireless scenarios [170]. For the heterogeneous scenarios with wired/wireless communications and dynamic on/off-line clients, how to maximize BC4LLM performance still needs to be researched.

### 6.5 Model interpretability

Model interpretability of AI training has long been a common topic in the field of deep neural networks. Although BC4LLM uses blockchain technology to ensure the whole process security of LLM, including learning corpus, training process, and generated content, model interpretability has not yet been solved [70].

Some studies have shown that when the parameter scale of an AI model is increased to a critical size (e.g. 10 B), some capabilities will leap forward in unexpected ways [171-172], such as contextual learning and stepwise reasoning. Such phenomena are fascinating, but also baffling. When and how LLMs acquire these capabilities is unclear. Recent research has either researched the factors contributing to this ability through a large number of experiments [173-175] or discussed some specific capabilities using existing theories [176-177]. However, more formal theories to explain such phenomena of LLMs are still lacking. These basic problems are worth studying and are of great significance for developing the next-generation LLMs and BC4LLM.

## 7. Conclusion

This paper proposes a blockchain-enabled LLM, namely BC4LLM, which can realize the whole process security of LLM through the security attributes of blockchain, including reliable learning corpus, secure training process, and identifiable generated content. It provides the realization path for the trusted AI.

In our paper, we introduce the background knowledge of blockchain, LLM, and blockchain-based AI. Then, we illustrate the security requirements of LLM for blockchain and how blockchain can transform LLM, i.e., how BC4LLM is designed. Moreover, we go further and give the potential application areas of BC4LLM and explain how it can play a role in these areas, especially in frontier communication network scenarios, which are network resource allocation, dynamic spectrum sharing, and semantic communication. Finally, we list five technical challenges of BC4LLM for researchers to consider further.

Today, the wave of LLMs is flooding human society with ChatGPT, and it is also changing the way people learn and work. This phenomenon will continue to deepen the concern about LLM-related technologies, and their security problems. The BC4LLM concept and technical route proposed in this paper provide reference and guidance for the realization of trusted AI, and provide inspiration for the new-generation LLM design.